\documentclass[11pt]{article}
\pdfoutput=1
\usepackage{fullpage}
\usepackage{cite}
\usepackage{graphicx}
\usepackage{amsmath}
\usepackage{amssymb}
\usepackage{subcaption}
\usepackage[linktocpage]{hyperref}
\title{}
\date{}
\renewcommand{\vec}[1]{{\bf #1} }

\newcommand{\nn}{\nonumber}
\def\beq{\begin{equation}}
\def\eeq{\end{equation}}

\begin{document}
\bibliographystyle{utphys}
\newcommand{\msbar}{\ensuremath{\overline{\text{MS}}}}
\newcommand{\DIS}{\ensuremath{\text{DIS}}}
\newcommand{\abar}{\ensuremath{\bar{\alpha}_S}}
\newcommand{\bb}{\ensuremath{\bar{\beta}_0}}
\newcommand{\rc}{\ensuremath{r_{\text{cut}}}}
\newcommand{\Nd}{\ensuremath{N_{\text{d.o.f.}}}}
\def\Ell{ {(L)} }
\def\z {\bar\zeta}

\titlepage
\begin{flushright}
BOW-PH-167\\     
NORDITA 2019-076\\
QMUL-PH-19-20\\
CERN-TH-2019-133

\end{flushright}

\vspace*{0.3cm}

\begin{center}
{\bf \Large A tale of two exponentiations in ${\cal N}=8$ supergravity}

\vspace*{1cm}

 \textsc{
Paolo Di Vecchia$^{a,b}$\footnote{divecchi@nbi.dk},
Andr\'{e}s Luna$^c$\footnote{luna@physics.ucla.edu}, 
Stephen G. Naculich$^d$\footnote{naculich@bowdoin.edu},
Rodolfo Russo$^e$\footnote{r.russo@qmul.ac.uk},
Gabriele Veneziano$^{f,g}$\footnote{gabriele.veneziano@cern.ch}
 and Chris D. White$^e$\footnote{christopher.white@qmul.ac.uk}} \\

\vspace*{0.8cm} $^a$ NORDITA, KTH Royal Institute of Technology and
Stockholm University, \\ 
Roslagstullsbacken 23, SE-10691 Stockholm, Sweden\\

\vspace*{0.2cm} $^b$ The Niels Bohr Institute, University of
Copenhagen, Blegdamsvej 17,\\
DK-2100 Copenhagen, Denmark\\

\vspace*{0.2cm} $^c$ Mani L. Bhaumik Institute for Theoretical
Physics, Department of Physics and Astronomy, University of California
at Los Angeles, California 90095\\

\vspace*{0.2cm} $^d$ Department of Physics, Bowdoin College, 
Brunswick, ME 04011 USA\\

\vspace*{0.2cm} $^e$ Centre for Research in String Theory, School of
Physics and Astronomy, \\
Queen Mary University of London, 327 Mile End
Road, London E1 4NS, UK\\

\vspace*{0.2cm} $^f$ Theory Department, CERN, CH-1211 Geneva 23, Switzerland\\

\vspace*{0.2cm} $^g$ Coll\'{e}ge de France, 11 place M. Berthelot,
75005 Paris, France\\

\end{center}

\vspace*{0.5cm}

\begin{abstract}
The structure of scattering amplitudes in supergravity theories
continues to be of interest. Recently, the amplitude for $2\rightarrow
2$ scattering in ${\cal N}=8$ supergravity was presented at three-loop
order for the first time. The result can be written in terms of an
exponentiated one-loop contribution, modulo a {\it remainder function}
which is free of infrared singularities, but contains leading terms in
the high energy Regge limit. We explain the origin of these terms from a
well-known, unitarity-restoring exponentiation of the high-energy 
gravitational $S$-matrix in impact-parameter space.
 Furthermore, we
predict the existence of similar terms in the remainder function at
all higher loop orders.  Our results provide a non-trivial
cross-check of the recent three-loop calculation, and a necessary
consistency constraint for any future calculation at higher loops.
\end{abstract}

\vspace*{0.5cm}

\section{Introduction}
\setcounter{equation}{0}
\label{sec:intro}

Scattering amplitudes in gauge and gravity theories continue to be
intensively studied, due to a wide variety of both formal and
phenomenological applications. Our focus in this paper is 
${\cal N}=8$
supergravity in four spacetime dimensions, which is of interest for a
number of reasons. Firstly, it may prove to be an ultraviolet finite
theory of perturbative quantum
gravity~\cite{Bern:2006kd,Bern:2007hh,Bern:2009kd,Bern:2014sna,Bern:2018jmv},
and in any case has a special status as its amplitudes arise in the
low energy limit from type II superstring theory~\cite{Green:1982sw}.
Secondly, calculations in
maximally supersymmetric theories can be simpler than in less
symmetric scenarios, making such theories the ideal frontier for
developing new calculational techniques. Thirdly, there are a number
of conjectures regarding the structure of amplitudes in maximally
supersymmetric theories, 
which higher-order computations are able to
shed light on.

One of the simplest amplitudes in terms of external multiplicity is
that of four-graviton scattering, results for which have been
previously calculated at
one-loop~\cite{Green:1982sw,Dunbar:1994bn,Dunbar:1995ed,Bern:2011rj}
and two-loop~\cite{Bern:1998ug,Naculich:2008ew,Brandhuber:2008tf,BoucherVeronneau:2011qv} order. In the maximally
supersymmetric theory, the tree-level result factors out, such that
the amplitude may be written in the form
\begin{equation}
i {\cal M}_4 = i{\cal M}_4^{(0)} 
\left( 1 +  \sum_{L=1}^\infty M_4^{(L)} \right),
\label{sugra-amp}
\end{equation}
where ${\cal M}^{(0)}$ is the tree-level contribution, and $M^{(L)}$
an implicitly defined correction factor at $L$-loop order. The latter
is infrared divergent, such that $M^{(L)}$ has a leading
$1/\epsilon^L$ pole in $d=4-2\epsilon$ spacetime dimensions.
Additional structure arises, however, from the fact that infrared
divergences in gravity theories are known to
exponentiate~\cite{Weinberg:1965nx,Naculich:2008ew,Naculich:2011ry,White:2011yy,Akhoury:2011kq,Naculich:2013xa,Melville:2013qca},
where the logarithm of the soft (IR-divergent) part of the amplitude
terminates at one-loop order, in marked contrast to (non-Abelian)
gauge
theories~\cite{Gatheral:1983cz,Frenkel:1984pz,Laenen:2008gt,Mitov:2010rp,Gardi:2010rn,Dixon:2008gr,Gardi:2011yz,Gardi:2013ita,Falcioni:2014pka,Vladimirov:2015fea}. This
motivates the following ansatz for the all-order amplitude:
\begin{equation}
i{\cal M}_4=i{\cal M}_4^{(0)}\exp[ M_4^{(1)}]\, {\cal F}_4,
\label{logA}
\end{equation}
where $M^{(1)}_4$ is the full one-loop correction factor, including
also its infrared singular part, and ${\cal F}_4$ is an infrared finite
{\it remainder function}, commencing at two-loop order. Indeed,
results for the latter have been presented at two-loop order for a
variety of supergravity theories in
ref.~\cite{Naculich:2008ew,Brandhuber:2008tf,BoucherVeronneau:2011qv}, 
and their implications discussed
further in refs.~\cite{Bartels:2012ra,Melville:2013qca}.

Recently, the four-graviton scattering amplitude in ${\cal N}=8$
supergravity has been obtained at an impressive three-loop
order~\cite{Henn:2019rgj}. The authors compared their results with the
form of eq.~(\ref{logA}), confirming that the three-loop remainder
function is infrared finite. This itself provided a highly non-trivial
cross-check of their results. However, as in previous
studies~\cite{BoucherVeronneau:2011qv,Bartels:2012ra,Melville:2013qca},
they also examined the behaviour of the remainder function in the high
energy {\it Regge limit}. This corresponds to highly forward high energy
scattering, such that the centre of mass energy is much greater than
the momentum transfer. The authors of ref.~\cite{Henn:2019rgj} noted
in particular the curious property that the remainder function,
although infrared finite, contains leading contributions in the high
energy limit, suggesting that their structure can be explained using
known results regarding high energy and / or soft limits. Indeed this
is the case, as we will show in this paper.

High energy scattering in gauge and gravity theories has been studied
for many decades. For example, generic scattering behaviour in the
Regge limit formed a crucial ingredient in the S-matrix programme of
strong interactions, which predated the discovery of QCD (see
e.g.~\cite{Eden:1966dnq} for a review). Obtaining similar behaviour in
perturbative quantum field theory has been pursued over many decades,
with relevant work in (super-)gravity
including~\cite{Lipatov:1982it,Lipatov:1982vv,Schnitzer:2007kh,Grisaru:1973vw,Grisaru:1973ku,Grisaru:1974cf,Grisaru:1981ra,Grisaru:1982bi,Naculich:2007ub,Schnitzer:2007rn}.
Recently,
methods from gauge theory have been used to analyse gravitational
physics, including clarifying the relationships between both theories
in certain kinematic
limits~\cite{Oxburgh:2012zr,Naculich:2011ry,Akhoury:2011kq,Melville:2013qca,Luna:2016idw,Saotome:2012vy,Vera:2012ds,Johansson:2013nsa,Johansson:2013aca,White:2014qia}.
Of particular relevance here is the outcome of the studies, 
started in the late 1980's~\cite{ Amati:1987wq, Muzinich:1987in, tHooft:1987vrq, Amati:1987uf, Amati:1990xe,Verlinde:1991iu,Amati:1992zb,Amati:1993tb,Giddings:2010pp,Kabat:1992tb,D'Appollonio:2010ae,Akhoury:2013yua,Collado:2018isu,KoemansCollado:2019lnh,KoemansCollado:2019ggb},
 of high-energy (transplanckian) gravitational scattering in the Regge-asymptotics regime\footnote{A different high-energy regime, at fixed scattering angle, was also considered at about the same time within string theory~\cite{Gross:1987ar,Mende:1989wt}.}
in both string and field theories (see \cite{Veneziano:2015kfb} for a recent review). 
Indeed, in order to explain the three-loop findings of ref.~\cite{Henn:2019rgj}, we
will use a very well-established property of gravitational scattering
in the leading Regge limit, namely that the
S-matrix has a
certain exponential structure in transverse position (i.e. impact parameter) space, in terms of the
so-called {\it eikonal phase}. This may be expanded in the
gravitational coupling constant, before being Fourier transformed to
momentum-transfer space order-by-order in perturbation theory. Given that a
product in position space\footnote{In the following we will often  refer to {\it transverse} space (momentum) as, simply, space (momentum), but it is important to stress that longitudinal momentum (energy) are never converted into the corresponding space (time) variables. This distinction is also important \cite{Amati:1987wq, Muzinich:1987in} to recover classical General Relativity expectations from the eikonal approximation when the eikonal phase is parametrically large.} is not a product (but rather a convolution)
in momentum space, the exponentiated eikonal phase in the former does
not directly lead to an exponential form in momentum space. The upshot
of this is that by making the ansatz of eq.~(\ref{logA}) in momentum
space, a mismatch occurs, giving leading Regge contributions in
the remainder function.

We will explicitly verify the form of the two- and three-loop
remainder functions in the (leading) Regge limit. Furthermore, we will
use our findings to predict additional terms at higher loops, before
forming a conjecture for the leading Regge behaviour of the remainder
function at arbitrary order in perturbation theory. Our results
provide a cross-check of the three-loop calculation in
ref.~\cite{Henn:2019rgj}, whilst also setting consistency
constraints on any future higher-loop calculations.

The structure of our paper is as follows. In section~\ref{sec:review},
we review previous results about fixed order results for supergravity
amplitudes, and also the exponentiation of the position space
amplitude in terms of the eikonal phase. In
section~\ref{sec:3loop}, we verify the form of the remainder
function up to three-loop order in the leading Regge limit. In
section~\ref{sec:nloop}, we extend our analysis to arbitrary orders in
perturbation theory.  Finally, in section~\ref{sec:discuss} we discuss our
results and conclude. 

\section{Review of previous results}
\label{sec:review}
\setcounter{equation}{0}

\subsection{The remainder function up to three-loop order}
\label{sec:remainder}

\begin{figure}
\begin{center}
\scalebox{0.5}{\includegraphics{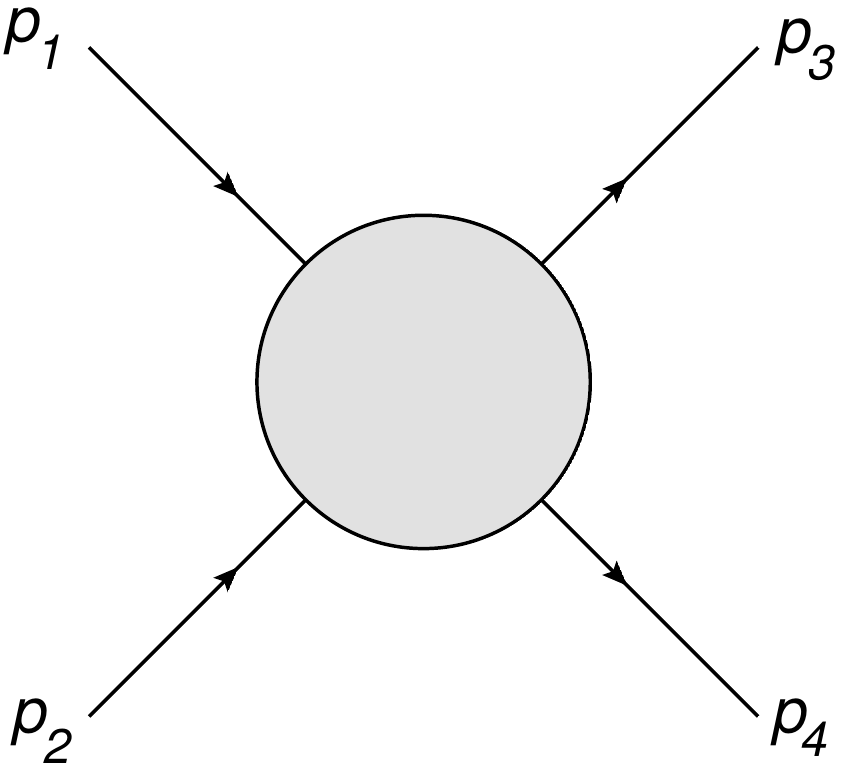}}
\caption{Labelling of 4-momenta for the four-graviton scattering
  process.}
\label{fig:2to2}
\end{center}
\end{figure}

As discussed above, the remainder function ${\cal F}_4$ of
eq.~(\ref{logA}) is defined after subtracting the one-loop
contribution from the logarithm of the 4-graviton scattering
amplitude. It thus begins at two-loop order, and we may then
consider the perturbative expansion
\begin{equation}
{\cal F}_4=1+\sum_{L=2}^\infty {\cal F}^{(L)}_4,
\label{Fexpand}
\end{equation}
where ${\cal F}_4^{(L)}$ is the $L$-loop contribution, including
coupling factors other than those associated with the tree-level
amplitude. Explicit results for the two-loop contribution (in a
variety of supergravity theories) have been presented in
ref.~\cite{Naculich:2008ew,Brandhuber:2008tf,BoucherVeronneau:2011qv}. 
To present results, we label
4-momenta as shown in figure~\ref{fig:2to2}, from which we may define
the Mandelstam invariants
\begin{equation}
s=(p_1+p_2)^2,\qquad t=(p_1-p_3)^2,\qquad u=(p_1-p_4)^2.
\label{Mandies}
\end{equation}
Note that all 4-momenta in figure~\ref{fig:2to2} are physical
(e.g. rather than all outgoing), so that we are dealing with the
physical scattering region
\begin{equation}
s\geq 0,\quad t,u\leq 0.
\label{physicalregion}
\end{equation}
Furthermore, momentum conservation implies 
$ s+t+u=0 $,
so that only two Mandelstam invariants are independent. The Regge
limit may then be formally defined as $s\gg -t$. Alternatively,
defining the dimensionless ratio
\begin{equation}
x=\frac{-t}{s},
\label{xdef}
\end{equation}
the Regge limit corresponds simply to $x\rightarrow 0$. 
Until recently, only the ${\cal O}(\epsilon^0)$ contribution
of the two-loop remainder
function was known, whose leading behaviour in the Regge limit may be
written as~\cite{Melville:2013qca}
\begin{align}
{\cal F}_4^{(2)}=x \left(\frac{\alpha_G s}{2} \right)^2 
&\bigg\{-2\pi^2\log^2 x-4\pi^2\log x +\pi^4+4\pi^2
\label{m48}\\
&
 +i\pi\left[\frac{4}{3}\log^3 x + 4\log^2x
+8\left(1+\frac{\pi^2}{3}\right)(1-\log x) + 16\zeta_3\right]\bigg\}+{\cal O}(x^2)+{\cal O}(\epsilon),
\nn
\end{align}
where we introduced the parameter 
\begin{equation}
\alpha_G \equiv  \frac{G_N}{\pi} (4\pi)^\epsilon \frac{ \Gamma^2
  (1-\epsilon) \Gamma (1+\epsilon)}{\Gamma (1-2\epsilon)} =
\frac{G_N}{\pi} +{\cal O}(\epsilon).
\end{equation}
Given that ${\cal F}_4^{(2)}$ is ${\cal O}(x)$, it vanishes in the strict Regge
limit. However, the results of ref.~\cite{Henn:2019rgj} have now
demonstrated that this is not true at higher orders in the dimensional
regularisation parameter $\epsilon$, nor at higher-loop level. 
In fact, the result in eq.~(6.5) of ref.~\cite{Henn:2019rgj}
is\footnote{The $\epsilon^2$ contribution to ${\cal F}_4^{(2)}$ is not
  explicitly written in~(6.5) of~\cite{Henn:2019rgj}, but can be
  deduced from the ancillary files attached to the arxiv version
  of~\cite{Henn:2019rgj}. The apparent sign discrepancy between
  ${\cal F}^{(3)}$ and eq.~(6.5)
results from our choosing $s>0$ whereas they have $s<0$.\label{sign}}
\begin{align}
{\cal F}^{(2)}_4&= \alpha_G^2 s^2 \pi^2 \left[3 \zeta_3\epsilon+ \left(\frac{\pi^4}{20}-  6 \zeta_3 \log(-t) \right) \epsilon^2 +{\cal O}(x)+{\cal O}(\epsilon^3)\right] \,,
\notag\\
{\cal F}^{(3)}_4&=-\frac{2}{3}i \alpha_G^3 s^3 \pi^3\zeta_3+{\cal O}(x)
+{\cal O}(\epsilon).
\label{Fresults}
\end{align}
These contributions are non-vanishing as $x\rightarrow 0$; we will
explain their origin in the following sections.

\subsection{Impact-parameter exponentiation and the eikonal phase}
\label{sec:eikphase}

\begin{figure}
\begin{center}
\scalebox{0.5}{\includegraphics{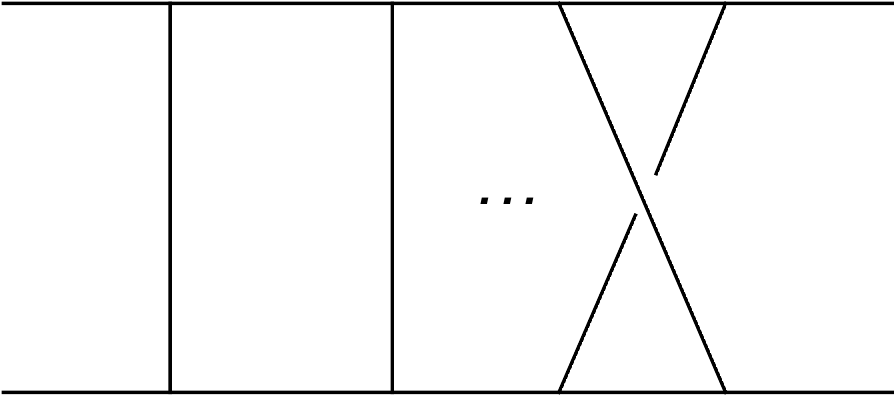}}
\caption{A representative (crossed) ladder graph, where all particles
  are gravitons. The sum of all such diagrams in the leading Regge
  limit builds up the exponentiated amplitude of eq.~(\ref{Meik}).}
\label{fig:ladders}
\end{center}
\end{figure}

The Regge limit of forward scattering consists of highly energetic
particles that barely glance off each other. As such, any exchanged
radiation must be soft (i.e. have an asymptotically small 4-momentum),
and the emitting particles are then said to be in the
 {\it eikonal approximation}. 
One may then show\cite{ Amati:1987wq, Muzinich:1987in} 
that the dominant behaviour at
arbitrary loop orders is given by the (crossed) horizontal ladder
graphs of figure~\ref{fig:ladders}, in which all particles are
gravitons. Furthermore, this situation does not depend on the amount
of supersymmetry: in the
leading Regge limit, the amplitude is dominated by the exchanged particle 
of highest spin, namely the graviton. 
It is then possible to sum such graphs to
all perturbative orders by working at fixed impact parameter $\vec{x}_\perp$, a $(d-2)$-dimensional vector
transverse to the incoming particle direction and which, at the leading eikonal level, can be thought as the transverse distance of closest approach between the two incoming hard gravitons. One
may then write the complete {\it eikonal amplitude}
as (see
e.g.~\cite{Levy:1969cr})
\begin{equation}
i{\cal M}_{\rm eik.}=2s\int d^{d-2}\vec{x}_\perp
e^{-i\vec{q}_\perp\cdot \vec{x}_\perp} \left(e^{i\chi(\vec{x}_\perp)}-1
\right),
\label{Meik}
\end{equation}
where the quantity $i\chi(\vec{x}_\perp)$ is known as the {\it eikonal
  phase}, and is given in $d= 4- 2\epsilon$ dimensions by\footnote{This result holds at finite $\epsilon$ and its validity is unrelated to the problem of infrared singularities originating in the $\epsilon \rightarrow 0$ limit discussed in this paper.} 
\cite{ Amati:1987wq, Muzinich:1987in}
\begin{equation}
i\chi(\vec{x}_\perp)=-iG_N s\Gamma(1-\epsilon)
\frac{(\pi\vec{x}_\perp^2)^\epsilon}{\epsilon}.
\label{chidef}
\end{equation}
In eq.~(\ref{Meik}), $\vec{q}_\perp$ is the $(d-2)$-dimensional
momentum transfer that is Fourier-conjugate to $\vec{x}_\perp$. In
terms of the above Mandelstam invariants, one has $t\simeq
-|\vec{q}_\perp|^2$ in the leading Regge limit. 
The exponentiation of the amplitude in terms of a large eikonal phase has the important consequence of restoring partial-wave unitarity, which is violated as $s \rightarrow \infty$ at  each loop order due to graviton exchange \cite{ Amati:1987wq, Muzinich:1987in}.
Equation~(\ref{Meik}) has a well-defined physical interpretation~\cite{Amati:1987wq, Muzinich:1987in, tHooft:1987vrq}, in
which $i\chi$ represents the phase shift experienced by one of the
incoming particles in the field set up by the other, thus forming a
link between old-fashioned quantum mechanical scattering theory and
perturbative QFT approaches (see e.g. ref.~\cite{Cheng:1987ga} for an
excellent review). Importantly, the exponentiation occurs in position
space. To obtain the momentum-space amplitude at a given order in
perturbation theory, one must Taylor expand the exponential in the
Newton constant $G_N$, before carrying out the Fourier
transform:
\begin{equation}
i{\cal M}_{\rm eik.}=2s \sum_{n=1}^\infty \frac{1}{n!}
\int d^{d-2}\vec{x}_\perp e^{-i\vec{q}_\perp\cdot
\vec{x}_\perp} [i\chi(\vec{x}_\perp)]^n.
\label{Meikexpand}
\end{equation}
In each term, the product of phase factors $[i\chi(\vec{x}_\perp)]^n$
becomes a convolution in momentum space, which may itself be given a
direct physical interpretation. First, one may express the
position-space eikonal phase as an inverse Fourier transform:
\begin{equation}
i\chi(\vec{x}_\perp)=-4\pi i G_N s\int\frac{d^{d-2}\vec{k}_\perp}
{(2\pi)^{d-2}}\frac{e^{i\vec{k}_\perp\cdot \vec{x}_\perp}}
{(-\vec{k}_\perp^2+i\varepsilon)},
\label{chimom}
\end{equation}
where $i\varepsilon$ denotes the usual Feynman prescription. This
allows us to rewrite eq.~(\ref{Meikexpand}) as
\begin{align}
i{\cal M}_{\rm eik.}&=2s \sum_{n=1}^\infty \frac{(-4\pi i G_N s)^n}{n!}
\int d^{d-2}\vec{x}_\perp e^{-i\vec{q}_\perp\cdot
\vec{x}_\perp}\left(\prod_{i=1}^n\int\frac{d^{d-2}\vec{k}_{i\perp}}
{(2\pi)^{d-2}}\frac{e^{i\vec{k}_i\cdot \vec{x}_\perp}}
{(-\vec{k}_{i\perp}^2+i\varepsilon)}\right)\notag\\
&=2s (2\pi)^{d-2}  \sum_{n=1}^\infty \frac{(-4\pi i G_N s)^n}{n!}
\prod_{i=1}^n\left(\int\frac{d^{d-2}\vec{k}_{i\perp}}
{(2\pi)^{d-2}}\frac{1}{(-\vec{k}_{i\perp}^2+i\varepsilon)}\right)
\delta^{(d-2)}\left(\vec{q}_\perp-\sum_{i=1}^n \vec{k}_{i\perp}\right).
\label{Meikexpand2}
\end{align}
Each term in the second line consists of a momentum space Feynman
integral, with $n$ particles being exchanged, each described by a
standard propagator in $(d-2)$-dimensions. It is the delta function
that makes this a convolution in momenta rather than a simple product,
and it simply corresponds to the fact that the sum of the exchanged
momenta should be equal to $\vec{q}_\perp$, namely the total momentum
transfer that is conjugate to the impact parameter. As we will see in
the following section, it is precisely the lack of a simple product in
momentum space that leads to the presence of the non-trivial remainder
function of eq.~(\ref{Fresults}).

\section{The three-loop remainder function in the Regge limit}
\label{sec:3loop}
\setcounter{equation}{0}

Having seen how to describe the leading Regge limit of the
four-graviton amplitude in supergravity to all orders via the eikonal
phase, we now have everything we need to explain the results of
ref.~\cite{Henn:2019rgj}, presented here in eq.~(\ref{Fresults}). To
obtain the $L$-loop remainder function, we may start with
eq.~(\ref{Meikexpand}), and identify $n=L+1$. Substituting
eq.~(\ref{chidef}) then yields
\begin{align}
i{\cal M}_{\rm eik.}&=2s\sum_{L=0}^\infty \frac{1}{(L+1)!}
\left(\frac{-i   G_N s\Gamma(1-\epsilon)\pi^\epsilon}{\epsilon}
\right)^{L+1}
\int d^{2-2\epsilon}\vec{x}_\perp e^{-i\vec{q}_\perp\cdot \vec{x}_\perp}
\left(\vec{x}_\perp^2\right)^{(L+1)\epsilon}\notag\\
&=2s\frac{4\pi i G_N s}{\vec{q}_\perp^2}
\sum_{L=0}^\infty \frac{1}{L!}\left[
-\, \frac{i G_N s\Gamma(1-\epsilon)}{\epsilon}
\left(\frac{4\pi}{\vec{q}_\perp^2}\right)^\epsilon\right]^L
\frac{\Gamma(1-\epsilon)\Gamma(1+L\epsilon)}
{\Gamma(1-(L+1)\epsilon)}.
\label{MeikL}
\end{align}
The second line allows us to identify the Regge limit of the
tree-level amplitude from the $L=0$ term:
\begin{equation}
  i {\cal M}_4^{(0)}=\frac{8\pi i G_N s^2 }{- t} 
+{\cal O}(x).
\label{M40eik}
\end{equation}
Examining the one-loop term then allows us to construct the correction
factor entering 
eq.~(\ref{logA})
\begin{equation}
M^{(1)}_4=\frac{i{\cal M}_4^{(1)}}{i{\cal M}_4^{(0)}}
=\, -\,\frac{i G_N s}{\epsilon}\frac{\Gamma^2(1-\epsilon)
\Gamma(1+\epsilon)}{\Gamma(1-2\epsilon)}
\left(\frac{4\pi}{-t}\right)^\epsilon.
\label{M14res}
\end{equation}
Let us now rewrite eq.~(\ref{MeikL}) as
\begin{align}
  i{\cal M}_{\rm eik.}&=i{\cal M}^{(0)}_4
\sum_{L=0}^\infty \frac{1}{L!}
\left[-\, \frac{i G_N s}{\epsilon}\frac{\Gamma^2(1-\epsilon)\Gamma(1+\epsilon)}
{\Gamma(1-2\epsilon)}\left(\frac{4\pi}{-t}\right)^\epsilon\right]^L
\notag\\ &\quad\qquad\qquad
\times
\left\{\frac{\Gamma^L(1-2\epsilon)\Gamma(1+L\epsilon)}
{\Gamma^{L-1}(1-\epsilon)\Gamma^L(1+\epsilon)\Gamma(1-(L+1)\epsilon)}
\right\}.
\label{MeikL2}
\end{align}
Were it not for the term in curly brackets, we would find that the
full momentum-space amplitude is simply the tree-level amplitude
multiplied by the {\it exponential of the one-loop correction} of
eq.~(\ref{M14res}). By comparing eqs.~(\ref{logA}) and~(\ref{MeikL2}),
we thus find that the remainder function is given by
\begin{align}
{\cal F}_4=\exp\bigl[-M_4^{(1)}\bigr]\sum_{L=0}^\infty
\frac{\bigl[M_4^{(1)}\bigr]^L}{L!}\left\{\frac{\Gamma^L(1-2\epsilon)\Gamma(1+L\epsilon)}
     {\Gamma^{L-1}(1-\epsilon)\Gamma^L(1+\epsilon)\Gamma(1-(L+1)\epsilon)}
     \right\}+{\cal O}(x).
\label{F4result}
\end{align}
This is a complete all-orders expression in the leading Regge limit
$x\rightarrow 0$, which may be systematically expanded in $G_N$ to
obtain the result at a given loop order. Performing such an expansion
(also in the dimensional regularisation parameter $\epsilon$), one
finds
\begin{align}
{\cal F}_4=1&+ \alpha^2_G s^2 \pi^2 
\left[ 3 \zeta_3\epsilon 
+ \left(\frac{\pi^4}{20}-  6 \zeta_3 \log(-t) \right) \epsilon^2 
+   {\cal O}(\epsilon^3)\right]
\nn\\
& 
+\alpha_G^3 s^3 \pi^3
\left[-\frac{2}{3}i\zeta_3
+{\cal O}(\epsilon^2)\right]
+{\cal O}(\alpha_G^4)+{\cal O}(x).
\end{align}
in agreement with eq.~(\ref{Fresults}) and thus
precisely confirming the results\footnote{See footnote \ref{sign}.}
of ref.~\cite{Henn:2019rgj}. 
We can now go further than this, however,
and predict the structure of the remainder function in the leading
Regge limit at higher orders in perturbation theory.

\section{The remainder function to all orders in the Regge limit}
\label{sec:nloop}
\setcounter{equation}{0}

In the previous section, we obtained a general expression,
eq.~(\ref{F4result}), for the remainder function ${\cal F}_4$ in the
leading Regge limit, and confirmed the results of a recent three-loop
calculation (which also necessarily included ${\cal O}(\epsilon)$ at
${\cal O}(G_N^2)$). However, the all-order nature of
eq.~(\ref{F4result}), in both $G_N$ and $\epsilon$, means that we can
expand this further. In doing so, we predict the existence of non-zero
terms in the remainder function at four-loop order and beyond. This
potentially provides a highly non-trivial cross-check of any future
calculations
in perturbative gravity.

We have expanded eq.~(\ref{F4result}) to 16 orders in $G_N$, finding
that all poles in $\epsilon$ vanish. This is to be expected, given the
aforementioned fact that all infrared singularities in gravity are
generated by the exponentiation of the one-loop
amplitude~\cite{Weinberg:1965nx,Naculich:2008ew,Naculich:2011ry,White:2011yy,Akhoury:2011kq,Naculich:2013xa,Melville:2013qca}.
Turning to the ${\cal O}(\epsilon^0)$ terms of the leading energy remainder 
${\cal F}_4 = {\cal F}_{4,0} + {\cal O}(\epsilon)+ {\cal O}(x)$, 
we may write the $L$-loop contribution as 
\begin{equation}
{\cal F}_{4,0}^{(L)}=(iG_Ns)^L f^{(L)},
\label{F4L}
\end{equation}
where we find the explicit results
\begin{align}
f^{(2)} &= 0 &
f^{(7)} &=  \z_7 &
f^{(12)} &=  \frac{1}{4!}\z_3^4 +  \z_9 \z_3+  \z_7 \z_5 \nn\\
f^{(3)} &=  \z_3 &
f^{(8)} &=  \z_5 \z_3  &
f^{(13)} &=  \frac12 \z_7 \z_3^2+ \frac12  \z_5^2 \z_3+ \z_{13} \nn\\
f^{(4)} &=    0 &
f^{(9)} &=   \frac{1}{3!} \z_3^3 + \z_9 &
f^{(14)} &=   \frac{1}{3!} \z_5 \z_3^3+ \z_{11} \z_3 +\frac12 \z_7^2 + \z_9\z_5
\label{finitecorrections} \\
f^{(5)} &=   \z_5 &
f^{(10)} &=    \frac12 \z_5^2+ \z_7 \z_3 &
f^{(15)} &=  \frac{1}{5!}  \z_3^5 +\frac12 \z_9 \z_3^2 + \z_7 \z_5 \z_3
+\frac{1}{3!} \z_5^3+ \z_{15} \nn\\
f^{(6)} &=   \frac12 \z_3^2 &
f^{(11)} &=    \frac12 \z_5 \z_3^2 + \z_{11} &
f^{(16)} &=   \frac{1}{3!} \z_7 \z_3^3 +\frac14 \z_5^2 \z_3^2 
+ \z_{13} \z_3+  \z_9 \z_7 +  \z_{11} \z_5 \nn
\end{align}
with $\z_n = 2 \zeta_n/n$. Despite the rather formidable nature of
eq.~(\ref{F4result}), we see that the results for the ${\cal
  O}(\epsilon^0)$ contributions have a simple form. It is apparent
that the arguments of the zeta functions in each term in the sums are
such that they form a partition of $L$ into a sum of odd integers
greater than one.  The generating function for the number of such
partitions is
\begin{equation}
\prod_{j=1}^\infty  \frac{1}{1-z^{2j+1} }
=
1+z^3+z^5+z^6+z^7+z^8+2 z^9+2 z^{10}+2 z^{11}+3 z^{12}+3 z^{13}+4 z^{14}
+5 z^{15}+5z^{16}+O\left(z^{17}\right) 
\label{partitions}
\end{equation}
so the coefficient of $z^L$ on the right-hand side of
eq.~(\ref{partitions}) tells us the number of individual terms in each
$f^{(L)}$ of eq.~(\ref{finitecorrections}). We then find that we can
summarise all of eq.~(\ref{finitecorrections}) as the compact formula
\begin{equation}
  {\cal F}_{4,0}^{(L)} = (i G_N s)^L \sum_{p_r(L)}
  \prod_{j} \frac{1}{n_j!} \left(\frac{ 2 \zeta_{L_j} }{L_j} \right)^{n_j},
\label{compact}
\end{equation}
where the sum is over all restricted partitions of $L$,
as mentioned above, the $L_j$'s are the distinct odd integers entering
in the partition and $n_j$ is the number of times each $L_j$ appears,
so we have
\begin{equation}
  L = \sum_j L_j n_j.
  \end{equation}
In fact, one may observe\footnote{We would like to thank Henrik Johansson for 
this observation.} 
that eqs.~\eqref{F4L}, \eqref{finitecorrections} and \eqref{compact} may be
compactly summarized by 
  \begin{align}
    \label{resum}
    {\cal F}_{4,0}
&= 1  + \sum_{L=2}^\infty {\cal F}^{(L)}_{4,0}
= \exp\left[ \sum_{j=1}^\infty (i G_N s)^{2j+1} \bar\zeta_{2j+1} \right] 
\nonumber \\ 
&  =  \frac{{\rm e}^{-2 i G_N s \gamma} }{\Gamma^2 (1+ i G_N s)} \exp \left[  \log\left( \frac{\pi  i G_N s}{\sin( \pi  i G_N s)}\right) \right] 
= {\rm e}^{-2i G_N s \gamma } \frac{\Gamma (1 - i G_N s)}{\Gamma (1+i G_N s)} \;.
  \end{align} 
The same result can be obtained from the $\epsilon\to 0$ limit of~\eqref{Meik} without expanding the exponential of $\chi(\vec{x}_\perp)$ (see ref.\cite{Melville:2013qca} for a similar observation). 
Denoting the $\epsilon^0$ terms of eqs.~\eqref{chidef} and~\eqref{M14res} 
by $\chi_0$ and $M_{4,0}^{(1)}$ respectively, we may write
  \begin{equation}
    \label{eq:chi0M40}
     \chi_0 = - G_N s \left( \log(\pi \vec{x}_\perp^2) + \gamma\right)~,\qquad\qquad
    {\rm e}^{M_{4,0}^{(1)}} = e^{i G_N s \gamma} \left( \frac{4\pi}{\vec{q}_\perp^2} \right)^{-i G_N s}.
  \end{equation}
We can then perform the Fourier transform of eq.~\eqref{Meik} in $d=4$ ({\rm i.e.} restricting to the $\epsilon$-independent part)
  \begin{align}
    \label{n2}
    \int d^2 \vec{x}_\perp {\rm e}^{-i\vec{q}_\perp\cdot \vec{x}_\perp} {\rm e}^{i\chi_0(\vec{x}_\perp)} = \frac{4\pi i G_N s}{\vec{q}_\perp^2} e^{-i G_N s \gamma} \left( \frac{4\pi}{\vec{q}_\perp^2}\right)^{-i G_N s} \frac{\Gamma (1 - i G_N s)}{\Gamma (1+i G_N s)} = \frac{4\pi i G_N s}{\vec{q}_\perp^2} e^{M_{4,0}^{(1)}} {\cal F}_{4,0},
  \end{align}
  and check explicitly that the last step is consistent with the result of eq.~\eqref{resum}\footnote{Note that, for $\epsilon \rightarrow 0$, the whole effect of ${\cal F}_{4,0}$ boils down to a renormalization of an unobservable (and not explicitly written) infinite Coulomb phase originating  from the leading eikonal resummation.}. This derivation can be seen as a proof of the result~\eqref{compact} for the $\epsilon^0$ contribution, but we
  stress in any case that a complete all-order expression for the
remainder function (which is more powerful than a finite-order
$\epsilon$ expansion) has already been given in eq.~(\ref{F4result}).

The next unknown order in the four-graviton amplitude is four
loops. It is easily checked from eq.~(\ref{F4result}) that, as at two
loops, the ${\cal O}(\epsilon^0)$ contribution to the remainder
function (in the leading Regge limit) vanishes. However, there is a
nonzero contribution beyond this, given by
\begin{equation}
{\cal F}_4^{(4)}=-5(G_N s)^4\zeta_5\epsilon
+{\cal O}(\epsilon^2) +{\cal O}(x).
\label{F4eps}
\end{equation}
We do not expect this result to be explicitly confirmed in the near
future: calculating the ${\cal O}(\epsilon)$ part of the four-loop
amplitude would presumably be first carried out as part of a five-loop
calculation!

An interesting observation is that the above results respect the
conjectured {\it uniform transcendentality} property of amplitudes in
theories with maximal supersymmetry. That is, we can associate a
transcendental weight $n$ with the zeta value $\zeta_n$, where all
rational coefficients are taken to have weight zero. The sum of
weights at ${\cal O}(\epsilon^m)$ and $L$-loop order is then
\begin{equation}
w=L+m.
\label{weight}
\end{equation}
Beyond the leading order, the Regge limit breaks this uniform
transcendentality property, as, for instance, one approximates
$\ln(-u/s)\sim x$ losing the transcendental contribution of the
logarithm.
Since the leading eikonal does not depend on the number of
supersymmetries, the uniform weight property for ${\cal N}=8$
supergravity manifest in~(\ref{weight}) is inherited by the lower
supersymmetric cases. We stress that this property of the leading term
is exact to all orders, not just the $\epsilon^0$ order considered
above. For the amplitude itself at a given loop order, there is a
dominant pole
\begin{displaymath}
\sim \frac{1}{\epsilon^L}
\end{displaymath}
coming from the exponentiated IR singularity in the one-loop
contribution. All subleading terms in $\epsilon$ (in the leading Regge
limit) come from expanding Euler gamma functions, and the coefficients
of all such expansions have increasing uniform weight as the power of
$\epsilon$ increases. Thus, this accounts precisely for the dependence
of eq.~(\ref{weight}).

\section{Discussion}
\label{sec:discuss}
\setcounter{equation}{0}

In this paper, we examined the form of the four-graviton scattering
amplitude in ${\cal N}=8$ supergravity, which was recently calculated
at three-loop order~\cite{Henn:2019rgj}. It is conventional to define
a remainder function for this amplitude, constituting what is left
upon factoring out the tree-level amplitude, and the one-loop
correction factor~\cite{BoucherVeronneau:2011qv}. The three-loop
calculation, which includes an evaluation of the ${\cal O}(\epsilon)$
part of the two-loop result, revealed the existence of leading terms
in the remainder function in Regge's  high energy limit, at
non-negative powers of the dimensional regularisation parameter
$\epsilon$. 

In this paper, we have shown that these contributions follow precisely
from the known exponentiation of the four-graviton amplitude in
position space, in terms of the so-called {\it eikonal phase}. At a
given order in perturbation theory, a product of one-loop amplitudes
occurs, which becomes a convolution in momentum space, whose physical
interpretation is that the transverse momentum transfer (conjugate to
the impact parameter) must be democratically shared between the
exchanged gravitons at that order. This in turn means that the
amplitude does not straightforwardly exponentiate in momentum space,
and we have derived an all-orders expression -- in both the
gravitational coupling $G_N$ and dimensional regularisation parameter
$\epsilon$ -- for the remainder function in the Regge limit. As well as
confirming the results of ref.~\cite{Henn:2019rgj}, we also predict
explicit contributions at higher loop orders.
We obtained a 
particularly convenient combinatorial form for the ${\cal O}(\epsilon^0)$ 
contributions, which we showed can be directly obtained from the
leading eikonal expression in $d=4$.
The higher loop remainder
function respects maximal transcendentality to all orders. 

There are a number of possible extensions of our
analysis. Firstly, one could look at predicting the structure of
subleading terms in the Regge limit 
(see e.g.
refs.~\cite{Amati:1990xe,Collado:2018isu,KoemansCollado:2019lnh,KoemansCollado:2019ggb,Amati:1993tb,Akhoury:2013yua,Luna:2016idw,Bartels:2012ra,SabioVera:2019edr}
for previous work in this area). Secondly, it would be interesting to
extend the analysis discussed in this paper to higher loops by
starting from the integral expressions for the four- and five-loop
amplitudes of refs.~\cite{Bern:2012uf,Bern:2017ucb}. Finally one can study the remainder
function in theories with less than maximal supersymmetry. This is not
independent of the exploration of subleading eikonal
contributions. Indeed, the leading Regge behaviour would be expected
to be the same for less supersymmetric gravity theories, given that
this kinematic regime is dominated by the exchange only of leading
soft particles of highest spin (i.e. the graviton). Three-loop
calculations in non-maximal supergravity theories do not yet exist,
thus our results already provide a highly useful constraint.

\section*{Acknowledgments}
We thank Zvi Bern, Lance Dixon, Henrik Johansson,
Julio Parra-Martinez, Lorenzo Magnea,
and Cristian Vergu for useful conversations.  
AL is supported in part by the 
Department of Energy under Award Number DESC000993.  
The research of SGN is supported in part by the National Science
Foundation under Grant No.~PHY17-20202.  
CDW and RR are supported by the UK Science and Technology 
Facilities Council (STFC) Consolidated Grant ST/P000754/1 
``String theory, gauge theory and duality, 
and/or by the European Union Horizon 2020 research and innovation programme
under the Marie Ck\l{}odowska-Curie grant agreement No. 764850 ``SAGEX''.  
AL and SGN wish to thank the Centre for Research in String Theory 
at Queen Mary University of London for hospitality.  
PDV, RR and GV would like to thank the Galileo Galilei Institute for
hospitality during the workshop ``String theory from the worldsheet
perspective'' where they started discussing this topic. 
PDV was supported as a Simons GGI scientist from the Simons
Foundation grant 4036 341344 AL. The research of PDV is partially
supported by the Knut and Alice Wallenberg Foundation under grant KAW
2018.0116.

\bibliography{refs.bib}
\end{document}